\documentclass{ws-ijmpe}
\usepackage[super,compress]{cite}

\def\half{{1\over 2}}

\def\fourth{{1\over4}}

\def\Z{{\mathchoice {\hbox{$\sf\textstyle Z\kern-0.4em Z$}}
{\hbox{$\sf\textstyle Z\kern-0.4em Z$}}
{\hbox{$\sf\scriptstyle Z\kern-0.3em Z$}}
{\hbox{$\sf\scriptscriptstyle Z\kern-0.2em Z$}}}}

\def\square{\kern1pt\vbox{\hrule height 1.2pt\hbox{\vrule width 1.2pt
   \hskip 3pt\vbox{\vskip 6pt}\hskip 3pt\vrule width 0.6pt}
   \hrule height 0.6pt}\kern1pt}
      \def\boxop{{\raise-.25ex\hbox{\square}}}
\def\mn{{\mu\nu}}

\def\tr{{\rm tr}\,}

\def\non{\nonumber\\}

\def\be{\begin{equation}}
\def\ee{\end{equation}\noindent}
\def\bear{\begin{eqnarray}}
\def\ear{\end{eqnarray}\noindent}
\def\bec{\blue\begin{equation}}
\def\eec{\end{equation}\black\noindent}
\def\bearc{\blue\begin{eqnarray}}
\def\earc{\end{eqnarray}\black\noindent}
\def\benn{\begin{enumerate}}
\def\enn{\end{enumerate}}

\def\slash#1{#1\!\!\!\raise.15ex\hbox {/}}
\newcommand{\slD}{\,\raise.15ex\hbox{$/$}\kern-.27em\hbox{$\!\!\!D$}}
\newcommand{\slpartial}{\raise.15ex\hbox{$/$}\kern-.57em\hbox{$\partial$}}

\newcommand{\nc}{\newcommand}
\nc{\spa}[3]{\left\langle#1\,#3\right\rangle}
\nc{\spb}[3]{\left[#1\,#3\right]}
\nc{\ksl}{\not{\hbox{\kern-2.3pt $k$}}}
\nc{\hf}{\textstyle{1\over2}}
\nc{\pol}{\varepsilon}
\nc{\tq}{{\tilde q}}
\nc{\esl}{\not{\hbox{\kern-2.3pt $\pol$}}}

\def\veps{\varepsilon}

\def\kinb{{1\over 4}\dot x^2}

\def\epseps#1#2{\varepsilon_{#1}\cdot \varepsilon_{#2}}
\def\epsk#1#2{\varepsilon_{#1}\cdot k_{#2}}
\def\kk#1#2{k_{#1}\cdot k_{#2}}

\def\4piTD{{(4\pi T)}^{-{D\over 2}}}
\def\4piT4{{(4\pi T)}^{-2}}

\def\Tintm4{{\dps\int_{0}^{\infty}}{dT\over T}\,e^{-m^2T}
    {(4\pi T)}^{-2}}
\def\Tintm{{\dps\int_{0}^{\infty}}{dT\over T}\,e^{-m^2T}}

\newcommand{\slG}{{{\dot G}\!\!\!\! \raise.15ex\hbox {/}}}
\newcommand{\Gd}{{\dot G}}

\def\GBd12{{\dot G}_{B12}}

\newcommand{\no}{\noindent}
\def\non{\nonumber}
\def\dps{\displaystyle}

\begin{document}

\markboth{N. Ahmadiniaz \& C. Schubert}{QCD gluon vertices from the string-inspired formalism}

\catchline{}{}{}{}{}
%%%%%%%%%%%%%%%%%%%%%%%%%%%%%%%%%%%%%%%%%%%%%%%%%%%%%%%%%%%%%%%%%%%%

\title{\bf QCD GLUON VERTICES FROM THE STRING-INSPIRED FORMALISM
%\footnote{For the title,
%try not to use more than 3 lines. Typeset the title in 10 pt
%Times roman, uppercase and boldface.}
}

\author{Naser Ahmadiniaz
%\footnote{Typeset names in
%10~pt Times roman, uppercase. Use the footnote to indicate
%the present or permanent address of the author.}
}

\address{Center for Relativistic Laser Science, Institute for Basic Science\
Gwangju 61005, Korea\\
%\footnote{State completely without abbreviations, the
%affiliation and mailing address, including country. Typeset in 8~pt
%Times italic.}
Ahmadiniaz@ibs.re.kr}

\author{Christian Schubert}

\address{ Instituto de F{{\'\i}}sica y Matem\'aticas, Universidad Michoacana de San Nicol\'as de Hidalgo\\
Apdo. Postal 2-82, C.P. 58040, Morelia, Michoacan, Mexico\\
christianschubert137@gmail.com}

\maketitle

\begin{history}
%\received{Day Month Year}
%\revised{Day Month Year}
%\accepted{Day Month Year}
%\comby{(xxxxxxxxxx)}
\end{history}

\begin{abstract}
The Bern-Kosower formalism, developed around 1990 as a novel
way of obtaining QCD amplitudes as the limit of infinite string tension
of the corresponding string amplitudes, was originally designed as an on-shell
formalism. Building on early work by Strassler, the authors have recently shown 
that this ``string-inspired formalism'' is extremely efficient also as a tool
for the study of off-shell amplitudes in QCD, and in particular for achieving
compact form factor decompositions of the N-gluon vertices. Among other things,
this formalism allows one to achieve a manifestly gauge invariant decomposition of these vertices
by way of integration-by-parts, rather than the usual tedious analysis of the nonabelian
off-shell Ward identities, and to combine the spin zero, half and one cases.
Here, we will provide a summary of the method, as well as its application to the 
three- and four-gluon vertices. 
%We also shortly comment on the abelian case.
%The abstract should summarize the context, content
%and conclusions of the paper in less than 200 words. It should
%not contain any references or displayed equations. Typeset the
%abstract in 8~pt Times roman with baselineskip of 10~pt, making
%an indentation of 1.5 pica on the left and right margins.
\end{abstract}


\ccode{PACS numbers:  11.15.-q , 12.38.-t, 12.38.Bx.} 

%\tableofcontents

\section{Introduction}

Recent years have seen an explosive development in the area of the calculation of on-shell matrix elements in quantum field theory.
particularly for gauge theory and gravity.  A whole host of new concepts and techniques have emerged, such as 
unitarity-based methods \cite{bddk-unitarity,berhua}, twistors \cite{witten-twistors}, BCFW recursion \cite{brcafe,bcfw},
and Grassmannians \cite{acck,masski}; see \cite{elvhua-review} and \cite{dixon-reviewnew} for recent reviews.

This sharply contrasts with the off-shell case, whose study has seen no comparable progress. 
Off-shell amplitudes in quantum field theory carry information that is often difficult, or even impossible, to
retrieve from the on-shell S-matrix. To mention just a few examples, off-shell information is useful for the full exploitation of the renormalization group,
the infrared properties of QCD \cite{alhusc}, and the matching of perturbative information with lattice data (see, e.g., \cite{petiws}).
Having explicit results, or at least well-organized integral representations, for off-shell amplitudes can also be highly useful
for the construction of higher-loop amplitudes, either directly or through the solution of the Schwinger-Dyson equations. 

Beyond the simplest cases, off-shell amplitudes generally depend on a large numbers of Lorentz invariants, so that
usually there is little hope for an explicit closed-form evaluation. The challenge is then rather to obtain integral representations
that are amenable to numerical evaluation, and well-adapted to the symmetries of the amplitude. In gauge theory or gravity, 
an important part of this task is to find a tensor decomposition in the polarization indices well-organized with respect to the
off-shell Ward identities. 

At the three-point level, a first systematic investigation of this problem was undertaken by Ball and Chiu in 1980. In \cite{balchi1}
they studied the vertex functions of scalar and spinor QED, and derived tensor decompositions consistent with the Ward identities and
free of kinematic singularities. The coefficient functions were calculated at the one-loop order. In \cite{balchi2} the same authors then did
a completely analogous study of the three-gluon vertex, and, in particular, found for it a decomposition in terms of six tensor
structures, or ``form factors'',  A,B,C,F, H, and S, where the last one actually turned out to be absent at one-loop. Of the others only the tensors F and H are transversal.

Although the actual calculations of \cite{balchi1,balchi2} were at the one-loop level, the obtained tensor decompositions (``Ball-Chiu-decompositions'') 
are of a universal character; only the coefficient functions of the various tensor structures will change at higher loop orders in perturbation theory.
It is therefore of considerable importance to obtain such decompositions also for the higher-point gluon vertices, i.e., the $N$ - gluon QCD amplitudes
(recall that, in the off-shell case, it is the one-particle irreducible (`1PI') amplitudes that constitute the natural carriers of information in quantum field theory).

It turns out, however, that the traditional way of deriving such tensor decompositions in gauge theory, based on the explicit analysis of the 
off-shell Ward identities, becomes extremely cumbersome beyond the three-gluon case. Clearly some method is called for that would allow
one to perform such a construction without having to solve the Ward identities. 

In a series of papers \cite{} the present authors have developed such a method, based on earlier work by Bern and Kosower, and Strassler.
Its essential feature is, that gauge-invariant structures are generated through certain integration-by-parts (`IBP') at the parameter integral level.
Around 1990,  Bern and Kosower \cite{berkos:prl,berkos:npb362,berkos:npb379} developed their well-known ``Bern-Kosower formalism'',
which allowed them to derive a new set of ``Bern-Kosower rules'' for the construction of one-loop QCD amplitudes by an analysis of the field theory limit
of the corresponding amplitudes in string theory. An essential technical ingredient of those rules is an IBP algorithm
that has a homogenizing effect on the integrands appearing in this limit, and allows one to use the underlying worldsheet supersymmetry to relate
the contributions of different spins in the loop. 
 
This formalism was restricted to the on-shell case, but shortly afterwards Strassler \cite{strassler1} constructed
a similar formalism inside field theory, using the worldline path integral representation of the effective action \cite{} as a starting point. 
Since the effective action is just the generating function of the 1PI Green's functions, this formalism is naturally geared towards the study of those. 
In a remarkable but unpublished paper \cite{strassler2} Strassler then showed, that the IBP algorithm used in the Bern-Kosower formalism
also leads to the automatic appearance of gauge-invariant structures in the effective action {\it at the integrand level}.  
Specifically, gauge fields $A^{\mu}$ are found to assemble into nonabelian field strength tensors (see \ref{conv} for our conventions) 

\bear
F_{\mn} \equiv F_{\mn}^a T^a = (\partial_{\mu}A_{\nu}^a - \partial_{\nu}A_{\mu}^a) T^a + ig[A_{\mu}^bT^b,A_{\nu}^cT^c]
\label{defF}\,,
\ear
where the terms linear in $A^{\mu}$  appear in the bulk, and the commutator term arises as a boundary term in the IBP. 
Strassler considered only the low-energy limit of the effective action, and the corresponding low-energy limit of the 
one-loop gluon amplitudes. 

In \cite{92} we applied the same methods, with some improvements on the IBP procedure \cite{26,91},
to the three-gluon amplitudes at full momentum, and showed that indeed it provides an extremely simple and elegant way of
rederiving the Ball-Chiu decomposition, as well as its coefficient functions at one loop, without the use of the Ward identity.
And it allows one to relate the spin zero, half and one contributions to the amplitude by the same ``replacement rules'' that are
part of the Bern-Kosower rules. Moreover, at the three-point level there are already ambiguities in the IBP which can be used
to optimize the integrand either with respect to gauge invariance, or with respect to transversality. The first representation, called
``Q-representation'' \cite{91}, arising from the most straightforward way of carrying out the IBP procedure, 
allows for a direct match of the vertex with the effective action; the second one, called ``S-representation'',
is the one that matches with the Ball-Chiu decomposition, and is characterized by manifest transversality of the integrand in
the bulk, the whole non-transversality  of the one-loop vertex having been absorbed into the boundary terms appearing in the IBP.
In terms of the structures A, B, C, F, H this means that the transversal ones, F and H, arise from the bulk integrand, and the non-transversal
ones A, B, and C as boundary terms. 

In \cite{98,bigone} we applied this method to the four-gluon vertex, and found everything to work quite the same way as in the three-point case:
The IBP procedure leads, at the integrand level, to a decomposition of the 1PI amplitude in terms of 19 tensor structures, well-organized with
respect to gauge invariance, and unifying the spin 0, half, and one cases. Of these 19 form factors, only 14 are true four-point form factors
(in the sense that their coefficient functions are given by typical four-point tensor parameter integrals, depending on the full set of kinematic invariants), 
while the remaining five arise as boundary terms. In the four-point case, one has to already distinguish betweens single and double boundary terms, given by three-point and two-point
parameter integrals, respectively, and of those five form factors two are of the first, three of the second kind. The Q-representation can be matched
to the effective action, while the S-representation has the property that all bulk terms are manifestly transversal at the integrand level, so that
it can be considered as the four-point generalization of the Ball-Chiu decomposition. For the S-representation the five ``boundary
form factors'' are just the structures A, B, C, F, H again, now reappearing with ``pinched'' momenta, F and H arising as single and A, B, C as double boundary terms.

The purpose of the present article is to review and summarize these recent results. Its structure is the following:
In section \ref{vertices} 
we collect some generalities on the $N$-gluon vertices, and review previous work on the three-gluon and four-gluon
cases (little or nothing seems to have been done yet beyond the four-gluon case). In section \ref{sec:3gluon} 
we review the Ball-Chiu decomposition of the three-gluon vertex. In section \ref{sec:bk} we summarize the work of Bern and Kosower, and
in section \ref{sec:strassler} describe the approach based on worldline path integrals, on which our method is based. 
In section \ref{sec:IBP} we discuss the IBP procedure, which is the technical centerpiece of this formalism. 
In section \ref{sec:3gluonSI} we present our rederivation of the three-point Ball-Chiu decomposition, and
in section \ref{sec:4gluonSI} our novel representation of the four-gluon vertex. The resulting tensor decomposition of the
four-gluon vertex is given in section \ref{sec:Tlist}, and a list of the corresponding one-loop parameter integrals in section \ref{sec:Plist}. 
Our conclusions are given in section \ref{sec:conc}.

\section{The N-gluon vertices}
\label{vertices}

Recall, that in non-abelian gauge theory one has the Lagrangian

\bear
{\cal L} = -\fourth  F^a_{\mn}F^{a\mn}\,,
\ear
with the non-abelian field strength tensor (\ref{defF}). The terms quadratic in $A_{\mu}$ provide the kinetic term, while the terms involving the
commutator produce a three-gluon vertex 
\bear
V_{\mu_1\mu_2\mu_3}^{a_1a_2a_3} = -igf^{a_1a_2a_3}\bigl[g_{\mu_1\mu_2}(k_1-k_2)_{\mu_3} + {\rm cycl.} \bigr]\,,
\label{3gltree}
\ear
and a four-gluon vertex
\bear
V_{\mu_1\mu_2\mu_3\mu_4}^{a_1a_2a_3a_4} &=& - g^2
\bigl[
f^{a_1a_2e}f^{a_3a_4e}(g_{\mu_1\mu_3}g_{\mu_2\mu_4} - g_{\mu_1\mu_4}g_{\mu_2\mu_3}) \nonumber\\
&& \quad + f^{a_1a_3e}f^{a_4a_2e}(g_{\mu_1\mu_4}g_{\mu_3\mu_2} - g_{\mu_1\mu_2}g_{\mu_3\mu_4}) \nonumber\\
&& \quad + f^{a_1a_4e}f^{a_2a_3e}(g_{\mu_1\mu_2}g_{\mu_4\mu_3} - g_{\mu_1\mu_3}g_{\mu_4\mu_2})\bigr]\,.\nonumber\\
\label{4gltree}
\ear
These vertices get corrected at the one-loop level by more complicated tensor structures, but the multiplicative renormalizability predicts that 
the UV- divergent parts of these corrections must take the same form as these tree-level vertices. 
Since the tree-level vertices are tied up with the kinetic term by gauge invariance, it is clear that
also the role of those UV divergences in the three-gluon and four-gluon functions must just be
to covariantize the ones contained in the two-point function, the gluon propagator. 
We can thus anticipate that they must arise from integrals of the vacuum polarization type. 

We will denote the contribution to the one-loop $N$-gluon vertex due to a particle with spin $s$ in the loop,
and with the standard ordering of the gluons $(12\ldots N)$,  by
$\Gamma^{a_1a_2\cdots a_N}_{s\,\mu_1\ldots \mu_N}[\veps_1,k_1;\ldots,\veps_N;k_N]$. 
For $s=1$ this is understood to denote the sum of the gluon and ghost loop contributions. 
Although we will only consider the fully off-shell case, we will nonetheless introduce polarization
vectors $\veps_1,\ldots ,\veps_N$ to write
\bear
\Gamma_s^{a_1a_2\cdots a_N}[\veps_1,k_1;\ldots,\veps_N;k_N] \equiv \veps_1^{\mu_1}\cdots \veps_N^{\mu_N}\Gamma^{a_1a_2\cdots a_N}_{s\mu_1\ldots \mu_N}[k_1,\ldots,k_N]\,.
\ear
These polarization vectors are arbitrary, and just serve as a book-keeping device that will allow us to write
some tensors in a more compact way in terms of the field strength tensors for the individual gluons.

Since our method maintains the full permutation (bose) symmetry between the gluons, it will be sufficient to
consider the contribution with the standard ordering.
The full amplitude is given by summing over all inequivalent (non-cyclic) permutations:
\bear
\Gamma_s[\veps_1,k_1;\ldots,\veps_N;k_N]
= \sum_{\pi\in S_N/Z_N} \Gamma_s^{a_{\pi(1)}a_{\pi(2)}\cdots a_{\pi(N)}}[\veps_{\pi(1)},k_{\pi(1)};\ldots,\veps_{\pi(N)};k_{\pi(N)}]\,.
\nonumber\\
\label{permute}
\ear
The gluon loop contribution will depend on the gauge fixing. Our method uses a path integral
representation of the gluonic effective action \cite{strassler1,18} that is based on the background 
field method with quantum Feynman gauge. 
This gauge choice is known \cite{binbro} to unify the gluon-loop contributions to the $N$ - gluon
vertices with the ones from scalars and fermions in the loop, in the sense that 
they then obey the same Ward identities, namely  \cite{corpap,papavassiliou-4gluon,bigone}
\bear
k_1^{\mu_1}\Gamma_{s\,\mu_1\ldots \mu_N}^{a_1a_2\cdots a_N}[k_1,\ldots,k_N]
&=&-ig f_{a_1a_2c}\Gamma^{ca_3a_4\cdots a_N}_{s\,\mu_2\ldots \mu_N}[k_1+k_2,k_3,\cdots,k_N]
\nonumber\\
&& 
-ig f_{a_1a_3c}\Gamma^{a_2ca_4\cdots a_N}_{s\,\mu_2\ldots \mu_N}[k_1,k_2+k_3,\cdots,k_N] + \ldots \,.\nonumber\\
\label{ward}
\ear
These Ward identities are inhomogeneous, mapping the $N$ - gluon vertex to the $N-1$ -gluon one.
For a generic gauge fixing, the right-hand-side would be different for the gluon loop case, and also involve the ghosts.

After these generalities, let us now review what has already been done on the gluon vertices for
the three- and four-point cases.  

The one-loop contribution to the three-gluon vertex due to a gluon in general covariant gauge 
was studied in 1979 by Celmaster and Gonsalves \cite{celgon}, although only at the symmetric point.
Kim and Baker \cite{kimbak} calculated the ghost-ghost-gluon vertex and the gluon and ghost self-energies, and
used the Ward identity to get from this an expression for the longitudinal part of the three-gluon vertex. 
Shortly afterwards Ball and Chiu presented their above-mentioned work on the decomposition of the three-gluon 
vertex at general momentum \cite{balchi2}, where they also calculated the one-loop contribution due to a gluon in Feynman gauge. 
Cornwall and Papavassiliou \cite{corpap} in 1989 constructed a ``gauge invariant three-gluon vertex'', fulfilling the ghost-free
Ward identity (\ref{ward}), through the pinch technique (see \cite{binpap-rev} for a review of this technique).
Freedman et al. in 1992 studied the conformal properties of this vertex \cite{fgjr}.
Davydychev, Osland and Tarasov \cite{daosta-3gluonD} in 1996 
calculated the gluon loop contribution to the one-loop three-gluon vertex
in arbitrary covariant gauge, and also the massless fermion loop contribution.
The fermion loop calculation was later generalized to the massive case by 
Davydychev, Osland and Saks \cite{daossa-3gluonDm}. 
In the already mentioned work by Binger and Brodsky \cite{binbro} they
studied the one-loop three-gluon vertex in various dimensions, 
using the background field method \cite{abbott,abgrsc}. 
Here besides the gluon and fermion loop cases they also included the scalar loop,
as is needed for SUSY extensions of QCD. 
They derived various sum rules relevant
to the SUSY case. 

The study of the loop corrections to the four-gluon vertex also started around 1980 with the work of Pascual and Tarrach \cite{pasctarr-80} 
who studied the four-gluon vertex coupling constant renormalization based on Weinberg's renormalization scheme \cite{weinberg-73}. 
In 1986, Brandt and Frenkel \cite{branfren-86} studied the infrared behavior of the three- and four-gluon vertices in Yang-Mills theory. 
In their analysis the four-gluon vertex was considered with all external gluons on-shell and transverse in the Feynman gauge for simplicity. 
%The process in their interest was independent of the choice of gauge because it is a gauge invariant process. 
They found that the 1PI  contribution to the four-gluon vertex exhibits single- and double-pole singularities. 
Papavassiliou in 1993 \cite{papavassiliou-4gluon} generalized the pinch-technique approach of \cite{corpap} to the four-gluon case,
and showed that this vertex again fulfilled the ghost-free Ward identity (\ref{ward}).
In 2008, Kellermann and Fischer \cite{kelfis} studied the running coupling constant of the four-gluon vertex in Landau gauge for pure Yang Mills theory. 
They investigated the non-perturbative structure of the vertex using the Dyson-Schwinger (`DS') equations for several momentum configurations.
A good agreement between their analytical results for the leading infrared and ultraviolet terms of the DS equation and their numerical solution was obtained. 

So far no information on the full off-shell four-gluon vertex structure has been presented in previous studies. 
In 2014, Gracey \cite{gracey-14} studied this structure at the symmetric point. 
He constructed the tensor structure of the vertex in terms of 138 Lorentz tensors which is the number of rank 4 Lorentz tensors built from the three independent external momenta and the metric.
Based on the determination of the full structure of this vertex a new momentum subtraction scheme was defined.
Binosi et. al \cite{biibpa-14} studied the nonperturbative structure of the 1PI part of the four-gluon vertex in the Landau gauge.  
In particular, they considered a subset of diagrams corresponding to the one-loop dressed diagrams with vertices to be kept at tree-level and fully dressed propagators. 
Their analysis was based on a very simple momentum configuration $(p,p,p,-3p)$. The infrared behavior of the gluon propagator was studied and a nonperturbative connection between 
the masslessness of the ghost  and the shape of the gluon propagator found for certain kinematic limits. Based on this connection they also predict the same behavior for any purely gluonic $N$-point function. 
Within the mentioned class of diagrams and with their momentum configuration they only found two orthogonal Lorentz and color tensor structures, see \cite{biibpa-14} for more details.  

Recently, Eichmann et. al \cite{eifihe-15} also considered the tensor structures of four point functions with external gauge bosons using the permutation group $S_4$. 
For the off-shell four point functions they predict 136 tensor structures. 
A DS study of the four-gluon vertex of Landau gauge Yang Mills theory has been carried out recently \cite{cyhusm-15} based on these 136 tensorial structures.
Their method of solving the DS equations used a truncation that included only leading diagrams in the ultraviolet, and lower Green functions from previous DS calculations that are in good agreements with lattice data. 
The running coupling constant was also studied. 

At two loops, so far both the three-gluon vertex \cite{daosta-2loop,davosl,gracey} and the four-gluon vertex \cite{gracey-14} 
have been studied only for very special momentum configurations.

\section{The three-gluon vertex and its Ball-Chiu decomposition}
\label{sec:3gluon}

\begin{figure}[h]
%\begin{picture}(0,0)(6000,10000)
\centering
\includegraphics[width=2.5in]{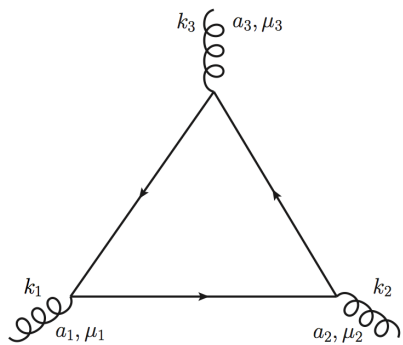}
\caption{Three-gluon vertex.}
\label{fig1}
\end{figure}

\no
The three-gluon vertex in QCD at tree level (\ref{3gltree})
is corrected at the  one-loop level by the 1PI  three-gluon vertex  with a spinor or gluon loop.
E.g. for the spinor loop case we have the diagram shown in Fig. \ref{fig1} (and a second one with the
other orientation of the fermion).

%The general (off-shell) three-gluon vertex, proposed by Ball and Chiu (1980)

\noindent
The Ball-Chiu decomposition of the vertex is \cite{balchi2}
\begin{eqnarray}
\Gamma_{\mu_{1}\mu_{2}\mu_{3}}(k_{1},k_{2},k_{3})&=&
A(k_{1}^2,k_{2}^2;k_{3}^2) g_{\mu_{1}\mu_{2}}(k_{1}-k_{2})_{\mu_{3}}
+ B(k_{1}^2,k_{2}^2;k_{3}^2)
g_{\mu_{1}\mu_{2}}(k_{1}+k_{2})_{\mu_{3}}\nonumber\\ 
&&\hspace{-70pt} 
+ C(k_{1}^2,k_{2}^2;k_{3}^2)
 [k_{1\mu_{2}}k_{2\mu_{1}}-k_{1}\cdot k_{2}g_{\mu_{1}\mu_{2}}](k_{1}-k_{2})_{\mu_{3}}\nonumber\\
 &&\hspace{-70pt} 
+\frac{1}{3} S(k_{1}^2,k_{2}^2,k_{3}^2)(k_{1\mu_{3}}k_{2\mu_{1}}k_{3\mu_{2}}+k_{1\mu_{2}}k_{2\mu_{3}}k_{3\mu_{1}})
\nonumber\\ &&\hspace{-70pt}
+ F(k_{1}^2,k_{2}^2;k_{3}^2)
[k_{1\mu_{2}}k_{2\mu_{1}}-k_{1}\cdot k_{2}g_{\mu_{1}\mu_{2}}]
[k_{2\mu_{3}}k_{1}\cdot k_{3}-k_{1\mu_{3}}k_{2}\cdot k_{3}]\nonumber\\&&\hspace{-70pt} +
 H(k_{1}^2,k_{2}^2,k_{3}^2)
 \Bigl(-g_{\mu_{1}\mu_{2}}[k_{1\mu_{3}}k_{2}\cdot k_{3}-k_{2\mu_{3}}k_{1}\cdot k_{3}]
+\frac{1}{3}(k_{1\mu_{3}}k_{2\mu_{1}}k_{3\mu_{2}}-k_{1\mu_{2}}k_{2\mu_{3}}k_{3\mu_{1}})\Bigr)\nonumber\\
&&\hspace{-70pt} + \, [ \mbox{\rm cyclic permutations of }(k_{1},\mu_{1}), (k_{2},\mu_{2}),(k_{3},\mu_{3})]\,.\nonumber\\
\end{eqnarray}
This form factor decomposition is universal, that is, 
valid for the scalar, spinor and gluon loop, and also for higher loop corrections. 
Only the coefficient functions $A,B,C,F,H,S$ change. 
At tree level, $A=1$, the other functions vanish. Explicit calculation shows that
$S$ still vanishes at one-loop. 
The tensor structures multiplying $F,H$ are manifestly transversal.

\section{The Bern-Kosower formalism}
\label{sec:bk}

In 1991 Bern and Kosower in their seminal work derived, by an analysis of the infinite string limit of 
certain string amplitudes, the following Bern-Kosower master formula 
\cite{berkos:prl,berkos:npb362,berkos:npb379}:
\bear
 \Gamma_0^{a_{1}\dots a_{N}}[k_{1},\varepsilon_{1};\dots;k_{N},\varepsilon_{N}]
 &=&(-ig)^{N}\mbox{tr}(T^{a_{1}}\dots T^{a_{N}})
 %(2\pi)^{D}i\delta(\sum p_{i})
 \int_{0}^{\infty} dT(4\pi T)^{-D/2}e^{-m^2 T}\nonumber\\
 &&\hspace{-40pt} \times\int_{0}^{T}d\tau_{1}\int_0^{\tau_{1}}d\tau_2\dots\int_0^{\tau_{N-2}}d\tau_{N-1}
 \nonumber\\ &&\hspace{-40pt}\times
 \exp\Bigg\{\sum_{i,j=1}^N\left[\frac{1}{2}
  G_{Bij}k_{i}\cdot k_{j}
-i\dot{G}_{Bij}\varepsilon_{i}\cdot k_{j}+\frac{1}{2}\ddot{G}_{Bij}\varepsilon_{i}\cdot\varepsilon_{j}\right]\Bigg\}
\Biggl\vert_{\rm lin (\varepsilon_1 \ldots \varepsilon_N)} \,.
\nonumber\\
\label{bk}
\ear
As it stands, this is a parameter integral representation for the (color-ordered) 1PI $N$ - gluon amplitude  induced by a {scalar} loop, with momenta
$k_i$ and polarizations $\varepsilon_i$, in $D$ dimensions.
Here $m$ and $T$ are the loop mass and proper-time, $\tau_i$ fixes the location of the $i$th gluon,
and $G_{Bij}\equiv G_B(\tau_i,\tau_j)$ denotes the ``bosonic''  worldline Green's function, defined by
\bear
G_{Bij}&=& |\tau_i - \tau_j | - \frac{(\tau_i - \tau_j)^2}{T}\,,
\ear
and dots generally denote a derivative acting on the first variable. Explicitly,
\bear
 \dot G_B(\tau_1,\tau_2)&=& {\rm sign}(\tau_1 - \tau_2)
- 2 {{(\tau_1 - \tau_2)}\over T}\, ,\nonumber\\
\ddot G_B(\tau_1,\tau_2)
&=& 2 {\delta}(\tau_1 - \tau_2)
- {2\over T} \,.
\nonumber\\
\label{GB}
 \ear
 Note that this master formula has the full permutation symmetry between the $N$ - gluons. 
 Note also that, although it is off-shell, it does not contain the Lorentz invariants $k_i^2$ and $\varepsilon_i\cdot k_i$. 
 
In the Bern-Kosower formalism, this master formula serves as a generating
functional for the full on-shell $N$ - gluon amplitudes for the scalar, spinor and gluon loop,
through the use of the {\it Bern-Kosower rules}:

\begin{itemize}
\item
For fixed $N$, expand the generating exponential and take only the terms linear in all polarization vectors.

\item
Use suitable integrations-by-parts (IBPs) 
to remove all second derivatives  $\ddot G_{Bij}$.

\item
Apply two types of pattern-matching rules:
\begin{itemize}
\item
The ``tree replacement rules'' 
generate (from a field theory point of view) the contributions of the missing reducible diagrams.

\item
The ``loop replacement rules''  generate 
the integrands for the spinor and gluon loop from the one 
for the scalar loop.
\end{itemize}

\end{itemize}

\section{The worldline path integral formalism}
\label{sec:strassler}

%Footnotes should be numbered sequentially in superscript
%lowercase roman letters.\footnote{Footnotes should be
%typeset in 8~pt Times roman at the bottom of the page.}

Shortly after the work of Bern and Kosower, Strassler \cite{strassler1} 
rederived the master formula and the loop replacement rules
using worldline path integral representations of the gluonic effective actions.
E.g. for the scalar loop, 
\be 
\Gamma [A]
=
{\rm tr}
\Tintm
\int{\cal D}x(\tau)\,
{\cal P}
e^{-\int_0^T d\tau\Bigl(
\kinb
+ig\dot x\cdot A(x(\tau))
\Bigr)}\,,
\nonumber\\
\ee\no
where $A_{\mu}=A_{\mu}^aT^a$ and $\cal P$ denotes path ordering.
This also showed that the master formula and the loop replacement rules hold off-shell, 
which was not obvious from its original derivation.
Moreover, in \cite{strassler2} Strassler noted that the IBP generates automatically abelian field strength tensors

\bear
f_i^{\mu\nu} \equiv
k_i^{\mu}\varepsilon_i^{\nu}
- \varepsilon_i^{\mu}k_i^{\nu}
\label{deff}\,,
\ear
in the bulk, and
color commutators $ [T^{a_i},T^{a_j}]$ as boundary terms.
Those fit together to produce full nonabelian field strength tensors (\ref{defF})
in the low-energy effective action.
Thus we see the {\it emergence of gauge invariant tensor structures at the integrand level}.

However, the removal of all $\ddot G_{Bij}$ by IBP can be done in many ways,
and it is not obvious how to do it in a way that would preserve also the bose symmetry between the gluons
at the integrand level.
In \cite{strassler2} Strassler started to investigate this ambiguity at the four-point level,
but an algorithm valid for any $N$ and manifestly preserving the permutation
invariance was found only much later \cite{26,41}.
This algorithm still followed the objective of achieving a form of the $N$ - gluon vertex that,
in $x$ - space, would correspond to a manifestly covariant representation of the
nonabelian effective action. However, it turns out not be optimized from another point
of view, which is important, e.g., for the Schwinger-Dyson equations, namely it does
not lead to a clean separation of the vertices into transversal and longitudinal parts.
This remaining obstacle has been overcome only recently in \cite{91}, where we
give two IBP algorithms that work for arbitrary $N$ and
%, combined with the loop replacement rules,
 lead to explicit form-factor decompositions of the off-shell $N$ - gluon amplitudes: 

\begin{itemize}

\item
The first algorithm uses only local total derivative terms and leads to a 
representation that matches term-by-term with the low-energy effective action
(``Q-representation '').

\item
The second algorithm uses both local and nonlocal total derivative terms and 
leads to the transversality of all bulk terms at the integrand level (`` S-representation'').

\end{itemize}

The S-representation involves $N$ reference vectors $r_1,\ldots, r_N$ fulfilling
$r_i\cdot k_i\ne 0$ but arbitrary otherwise. In terms of these reference vectors, the
transition from the Q-representation to the S-representation can, for the bulk
(but not the boundary) terms  also be simply stated as the following shift of all
polarization vectors:

\bear
\varepsilon_i \longrightarrow \tilde \varepsilon_i \equiv \varepsilon_i - \frac{\varepsilon_i \cdot r_i}{k_i \cdot r_i} k_i
\, .
\label{epsshift}
\ear
This makes transversality, i.e. vanishing under $\varepsilon_i \to k_i$, manifest. 

\no
In \cite{92} we applied both algorithms to the three-point case and showed that, in particular,  
the second algorithm generates precisely the Ball-Chiu decomposition.
Very recently, we have carried out the same program also for the much more challenging four-gluon vertex \cite{bigone}.
We will now sketch these rather involved calculations as well as space permits.

\section{The integration-by-parts procedure}
\label{sec:IBP}

A full discussion of the IBP procedure would be too lengthy to be presented here. Thus we will show here only an
example at the three-point level, and refer the reader to \cite{91} for an exhaustive discussion. 
For $N=3$, the expansion of the Bern-Kosower master formula (\ref{bk}) yields
\bear
\Gamma_{0}^{a_{1}a_{2} a_{3}}[k_{1},\varepsilon_{1};k_{2},\varepsilon_{2};k_{3},\varepsilon_{3}]
 &=&(-ig)^{3}\mbox{tr}(T^{a_{1}} T^{a_{2}}T^{a_{3}})\int_{0}^{\infty} dT(4\pi T)^{-D/2}e^{-m^2 T}\nonumber\\
 &&\hspace{-70pt}  \times\int_{0}^{T}d\tau_{1}\int_0^{\tau_{1}}d\tau_{2}\, (-i)^3 P_3
\, e^{(G_{B12}k_{1}\cdot k_{2}+G_{B13}k_{1}\cdot k_{3}+G_{23}k_{2}\cdot k_{3})}\,,\nonumber\\
\label{expand3point}
\ear
where
\bear
P_3 &=&
\dot{G}_{B1i}\varepsilon_{1}\cdot k_{i}\dot{G}_{B2j}\varepsilon_{2}\cdot k_{j}\dot{G}_{B3k}\varepsilon_{3}\cdot k_{k}
-\ddot{G}_{B12} \varepsilon_{1}\cdot\varepsilon_{2}\dot{G}_{B3k}\varepsilon_{3}\cdot k_{k}
\nonumber\\
&& -\ddot{G}_{B13}\varepsilon_{1}\cdot\varepsilon_{3}\dot{G}_{B2j}\varepsilon_{2}\cdot k_{j}
-\ddot{G}_{B23}\varepsilon_{2}\cdot\varepsilon_{3}\dot{G}_{B1i}\varepsilon_{1}\cdot k_{i}\,,
\nonumber\\ 
\label{P3} 
\ear
and we have introduced the convention that  repeated indices $i,j,k,\ldots $ are to be summed from 1 to $N=3$. 
To remove, e.g., the term involving $\ddot G_{B12}\dot G_{B31}$ in the second term of $k_3$, we add the total derivative
\bear
-\frac{\partial}{\partial \tau_2}\Bigl(\dot{G}_{B12} \varepsilon_{1}\cdot\varepsilon_{2}\dot{G}_{B31}\varepsilon_{3}\cdot k_{1}
e^{(G_{B12}k_{1}\cdot k_{2}+G_{B13}k_{1}\cdot k_{3}+G_{23}k_{2}\cdot k_{3})}\Bigr) \, .
\label{add}
\ear
In the abelian case this total derivative term would be integrated over the full circle, and the result would be zero, since the worldline
Green's function $G_B(\tau_1,\tau_2)$ has the appropriate periodicity properties to make the two boundary terms
cancel. Here instead we find a nonzero result:
\bear
-\dot{G}_{B12} \varepsilon_{1}\cdot\varepsilon_{2}\dot{G}_{B31}\varepsilon_{3}\cdot k_{1}e^{(\cdot)}
\Big\vert_{\tau_2=\tau_3}^{\tau_2=\tau_1}
= 0 + \dot{G}_{B13} \varepsilon_{1}\cdot\varepsilon_{2}\dot{G}_{B31}\varepsilon_{3}\cdot k_{1}e^{G_{B13}k_{1}\cdot (k_{2}+k_{3})}
\, .
\nonumber\\
\label{bt}
\ear
Now, in the three-point case there are already two inequivalent orderings, say, $(123)$ and $(132)$; thus the full amplitude will
also have a part $\Gamma^{a_{1}a_{3} a_{2}}$ with color trace $\tr (T^{a_1}T^{a_3}T^{a_2})$, and the same total derivative term will
contribute to it a boundary term
\bear
-\dot{G}_{B12} \varepsilon_{1}\cdot\varepsilon_{2}\dot{G}_{B31}\varepsilon_{3}\cdot k_{1}e^{(\cdot)}
\Big\vert_{\tau_2=\tau_1}^{\tau_2=\tau_3}
= - \dot{G}_{B13} \varepsilon_{1}\cdot\varepsilon_{2}\dot{G}_{B31}\varepsilon_{3}\cdot k_{1}e^{G_{B13}k_{1}\cdot (k_{2}+k_{3})} - 0 \, .
\nonumber\\
\label{btother}
\ear
These two boundary terms would cancel in the abelian case, but now instead combine to produce a color commutator $\tr (T^{a_1}[T^{a_2},T^{a_3}])$. 
Moreover, among the other five similar total derivative terms needed to remove all the $\ddot G_{Bij}$s from $k_3$ into $Q_3$ there is
one that differs from (\ref{add}) only by the interchange $2\leftrightarrow 3$. With some relabeling of integration variables, we can combine
the two boundary terms generated by that term with the two above to the structure
\bear
\tr (T^{a_1}[T^{a_2},T^{a_3}]) 
\varepsilon_3\cdot f_1\cdot\varepsilon_2
\dot{G}_{B12}\dot{G}_{B21}\,e^{G_{B12}k_{1}\cdot (k_{2}+k_{3})}  \, .
\label{structure}
\ear
This term involves only a two-point integral, with ``pinched'' momenta $k_2+k_3$, and it is easy to see that its
role is to provide a piece needed to extend the ``abelian'' Maxwell term $\tr(f_{\mn}f^{\mn})$ to the
full nonabelian one $\tr(F_{\mn}F^{\mn})$. 

\section{The three-gluon vertex in the string-inspired formalism}
\label{sec:3gluonSI}

We now present the final Q- and S- representations for the three-gluon vertex.

\subsection{The Q-representation of the three-gluon vertex}

For $N = 3$, the Q-representation is (for the scalar loop) \cite{92}
\bear
\Gamma &=& 
\frac{g^3}{(4\pi)^{\frac{D}{2}}}\mbox{tr}(T^{a_{1}} [T^{a_{2}},T^{a_{3}}])(\Gamma^{3} 
+ \Gamma^{2} + \Gamma^{{\rm bt}})\,,
\nonumber\\
\ear
where
\bear
\Gamma^{3} &=& - \int_{0}^{\infty} \frac{dT}{T^{\frac{D}{2}}}e^{-m^2 T}\int_{0}^{T}d\tau_{1}\int_0^{\tau_{1}}d\tau_{2}\, Q_3^3 
\exp\bigg\{\sum_{i,j=1}^3\frac{1}{2}G_{Bij}k_{i}\cdot k_{j} \bigg\}\,,
\nonumber\\
\Gamma^{2} &=& \Gamma^{3}(Q_3^3\to Q_3^2)\,, \nonumber\\
\Gamma^{{\rm bt}} &=&
\int_{0}^{\infty} \frac{dT}{T^{\frac{D}{2}}}e^{-m^2 T}\int_{0}^{T}d\tau_{1}
\dot{G}_{B12}\dot{G}_{B21} 
\Bigl\lbrack\varepsilon_3\cdot f_1\cdot\varepsilon_2
\,e^{G_{B12}k_{1}\cdot (k_{2}+k_{3})} +
{\rm cycl.}
\Bigr\rbrack\,, \nonumber\\
\ear
and
\bear
Q_{3}^3&=&\dot{G}_{B12}\dot{G}_{B23}\dot{G}_{B31}\tr(f_1f_2f_3)\,, \nonumber\\
Q_{3}^2&=& \half \dot{G}_{B12}\dot{G}_{B21}\tr (f_1f_2)\dot{G}_{B3k}\varepsilon_{3}\cdot k_{k}+
2 \, {\rm perm}\,.
\nonumber\\
\label{Q3}
\ear
Here $\Gamma^{\rm bt}$ comes from the boundary terms, and the upper indices on 
$\Gamma^{2,3},Q^{2,3}$ refer to the ``cycle content''; e.g. $Q_3^3$ contains a factor
$\dot{G}_{B12}\dot{G}_{B23}\dot{G}_{B31}$ whose indices form a closed cycle involving
three points, called ``three-cycle''.
To pass from the scalar to the spinor loop, one applies the ``loop replacement rules''
\bear
\dot G_{Bij}\dot G_{Bji} &\to & \dot G_{Bij}\dot G_{Bji} - G_{Fij}G_{Fji}\,, \nonumber\\
\dot G_{B12}\dot G_{B23}\dot G_{B31} &\to & \dot G_{B12}\dot G_{B23}\dot G_{B31} - G_{F12}G_{F23}G_{F31}\,, \nonumber\\
\label{spin}
\ear
where $G_{Fij} = {\rm sign}(\tau_i-\tau_j)$.
Similarly, the integrand for the gluon loop is obtained from the scalar loop one by
\bear
\dot G_{Bij}\dot G_{Bji} &\to & \dot G_{Bij}\dot G_{Bji} - 4G_{Fij}G_{Fji}\,, \nonumber\\
\dot G_{B12}\dot G_{B23}\dot G_{B31} &\to & \dot G_{B12}\dot G_{B23}\dot G_{B31} - 4G_{F12}G_{F23}G_{F31}\,. \nonumber\\
\label{gluon}
\ear
As stated above, the gluon loop vertex obtained in this way corresponds to the background field method with quantum Feynman gauge
\cite{strassler1,18}. 

%And for all three cases - scalar, spinor and gluon loop - the vertex allows a perfect match with the low-energy effective action.
%We recall that the low energy expansion of the one-loop QCD effective action induced by a
%loop particle of mass $m$ has the form (see, e.g., \cite{25})
%
%\begin{equation}
%\Gamma [F] = \int_0^\infty \!{dT\over T} \; 
%\frac{{\rm e}^{-m^2 T}}{(4\pi T)^{D/2}} \; 
%{\rm tr} \; \int \! dx_0 \; \sum_{n=2}^{\infty} \; 
%\frac{(-T)^n}{n!} \; O_n[F] \,,
%\nonumber
%\end{equation}\no
%where $O_n(F)$is a Lorentz and gauge invariant expression of mass dimension $2n$.
%To lowest orders,
%
%\begin{eqnarray}
%O_2 &=& c_2 g^2 F_{\mu\nu} F_{\mu\nu} \, ,\nonumber\\
%O_3 &=& 
%          c_3^3\,i g^3\,F_{\kappa\lambda}F_{\lambda\mu}F_{\mu\kappa} 
%          + c_3^2 g^2D_{\lambda}F_{\mu\nu}D^{\lambda}F^{\mu\nu} \,,
%          \nonumber\\
%\end{eqnarray}
%where only the coefficients $c_2,c_3^{2,3}$ depend on the spin of the loop particle. 
%We recognize the correspondences
%
%\begin{eqnarray}
%&&\Gamma^3\leftrightarrow F_{\kappa}^{~\lambda}F_{\lambda}^{~\mu}F_{\mu}^{~\kappa}
%=f_{\kappa}^\lambda f_{\lambda}^\mu f_{\mu}^\kappa+{\rm higher~point~terms}\nonumber\\
%&&\Gamma^2\leftrightarrow (\partial+ig\underbrace {A)F(\partial}+igA)F\nonumber\\
%&&\Gamma^{{\rm bt}}\leftrightarrow(f+ig\underbrace{[A,A])(f}+ig[A,A])\,.\nonumber\\
%\end{eqnarray}

\subsection{The S-representation of the three-gluon vertex}

\no
In the S-representation, the three-gluon vertex becomes
\bear
\tilde\Gamma &=& 
\frac{g^3}{(4\pi)^{\frac{D}{2}}}\mbox{tr}(T^{a_{1}} [T^{a_{2}},T^{a_{3}}])(\tilde\Gamma^{3} 
+ \tilde\Gamma^{2} + \tilde\Gamma^{{\rm bt}})\,,
\ear
where
\bear
\tilde\Gamma^{3} &=& - \int_{0}^{\infty} \frac{dT}{T^{\frac{D}{2}}}e^{-m^2 T}\int_{0}^{T}d\tau_{1}\int_0^{\tau_{1}}d\tau_{2}\, S_3^3 
\exp\bigg\{\sum_{i,j=1}^3\frac{1}{2}G_{Bij}k_{i}\cdot k_{j} \biggr \}\,,
\nonumber\\
\tilde\Gamma^{2} &=& \tilde\Gamma^{3}(S_3^3\to S_3^2)\,, \nonumber\\
\tilde\Gamma^{{\rm bt}} &=&
\int_{0}^{\infty} \frac{dT}{T^{\frac{D}{2}}}e^{-m^2 T}\int_{0}^{T}d\tau_{1}
\dot{G}_{B12}\dot{G}_{B21} 
\Bigl\lbrace\Bigl[ \varepsilon_3\cdot f_1\cdot\varepsilon_2
-\half {\rm tr}(f_1f_2)\rho_3+\half {\rm tr}(f_3f_1)\rho_2\Bigr]
\nonumber\\
&& \hspace{140pt}\times \,e^{G_{B12}k_{1}\cdot (k_{2}+k_{3})}  +\, {\rm cycl.}\Bigr\rbrace\,, \nonumber\\
\ear
and
\bear
S_{3}^3&=&\dot{G}_{B12}\dot{G}_{B23}\dot{G}_{B31}\tr(f_1f_2f_3)\,, \nonumber\\
S_{3}^2&=& \half \dot{G}_{B12}\dot{G}_{B21}\tr (f_1f_2)\dot{G}_{B3k}\frac{r_3\cdot f_3\cdot k_k}{r_3\cdot k_3}
+ 2 \, {\rm perm}\,.
\nonumber\\
\label{S3}
\ear
%\label{gammas0fin}
Here we have introduced three vectors $r_i$, and abbreviated $ \rho_i:=\frac{r_i\cdot\varepsilon_i}{r_i\cdot k_i}$.
Note that $S^3_3$ is the same as $Q^3_3$ above, but that in $S^2_3$, contrary to $Q^2_3$, all three polarization
vectors $\varepsilon_i$ are absorbed in abelian field strength tensors $f_i$.
Thus all bulk terms are now manifestly transversal, 
and it turns out that with the cyclic choice  $$r_{1}=k_2-k_3, r_{2}=k_3-k_1, r_{3}=k_1-k_2\,,$$ we 
get a term-by-term match with the Ball-Chiu decomposition:
\bear
H(k_1^2,k_2^2,k_3^2)&=&
C(r)\frac{g^2}{(4\pi)^{D/2}}\Gamma(3-\frac{D}{2}) I^{D}_{3,B} (k_1^{2},k_2^{2},k_3^{2})\,, \nonumber\\
A (k_1^2,k_2^2;k_3^2)&=&
C(r)\frac{g^2}{2(4\pi)^{D/2}}\Gamma(2-\frac{D}{2})\Bigl[I^{D}_{{\rm bt, B}}(k_1^2)+I^{D}_{{\rm bt, B}}(k_2^2)\Bigr]\,, \nonumber\\
B(k_1^2,k_2^2;k_3^2)&=&
C(r)\frac{g^2}{2(4\pi)^{D/2}}
\Gamma(2-\frac{D}{2})\Big[I^{D}_{{\rm bt, B}}(k_1^2)-I^{D}_{{\rm bt, B}}(k_2^2)\Big] \,,\nonumber\\
F(k_1^2,k_2^2;k_3^2)&=&
C(r)\frac{g^2}{(4\pi)^{D/2}}\Gamma(3-\frac{D}{2})
\frac{I_{2,B}^{D}(k_1^2,k_2^2,k_3^2)-I_{2,B}^{D}(k_2^2,k_1^2,k_3^2)}{k_1^2-k_2^2}\,, \nonumber\\
C(k_1^2,k_2^2;k_3^2)&=&
C(r)\frac{g^2}{(4\pi)^{D/2}}\Gamma(2-\frac{D}{2}) \frac{I^{D}_{{\rm bt, B}}(k_1^2)-I^{D}_{{\rm bt, B}}(k_2^2)}{k_1^2-k_2^2}\,,\nonumber\\
S(k_1^2,k_2^2;k_3^2)&=&0\,,\nonumber\\
\ear
where we have used  ${\rm tr}(T^{a_1}[T^{a_2},T^{a_3}])=iC( r)\,f^{a_1a_2a_3}$.
The coefficient functions appearing here are 
\bear
I_{3,B}^D(k_1^2,k_2^2,k_3^2) &=& \int_0^1d{\alpha}_1d{\alpha}_2d{\alpha}_3\delta(1-{\alpha}_1-{\alpha}_2-{\alpha}_3)
\times\frac{(1-2{\alpha}_1)(1-2{\alpha}_2)(1-2{\alpha}_3)}{\Bigl( m^2 + {\alpha}_1{\alpha}_2k_1^2+{\alpha}_2{\alpha}_3k_2^2+{\alpha}_1{\alpha}_3k_3^2\Bigr)^{3-\frac{D}{2}}}\,,\nonumber\\
I_{2,B}^D(k_1^2,k_2^2,k_3^2) &=& \int_0^1d{\alpha}_1d{\alpha}_2d{\alpha}_3\delta(1-{\alpha}_1-{\alpha}_2-{\alpha}_3)
\times\frac{(1-2{\alpha}_2)^2(1-2{\alpha}_1)}{\Bigl( m^2 + {\alpha}_1{\alpha}_2k_1^2+{\alpha}_2{\alpha}_3k_2^2+{\alpha}_1{\alpha}_3k_3^2\Bigr)^{3-\frac{D}{2}}}\,,\nonumber\\
I_{bt,B}^D(p^2) &=& \int_0^1d{\alpha}\frac{(1-2{\alpha})^2}{\bigl( m^2 + {\alpha}(1-{\alpha})k^2\bigr)^{2-\frac{D}{2}}}\,.\nonumber\\
\label{3pointints}
\ear
Here we have transformed from the integration variables $\tau_i$ to standard Feynman/Schwinger parameters $\alpha_i$ via
$\tau_{1,2} = Tu_{1,2}, \tau_3 = 0$ and 
\bear
\alpha_1&=&1-u_1, \quad
\alpha_2=u_1-u_2, \quad 
\alpha_3=u_2
 \label{tautoalpha}
\ear
with $\sum_{i=1}^3 \alpha_i=1$.

All this is written for the scalar loop case, but due to the loop replacement rules (which generalize (\ref{spin}) and (\ref{gluon})
in a straightforward way) the transition to the spinor and gluon loop cases is quite trivial, amounting only to simple algebraic changes of
the numerator  polynomials in (\ref{3pointints}).
This is, of course, very different from the standard Feynman diagram approach, where each spin in the loop
requires a separate calculation. 
\section{The four-gluon vertex in the string-inspired formalism}
\label{sec:4gluonSI}

\no
Proceeding to the four-point case, for the spinor loop case we have the diagram shown in Fig. \ref{fig2} (and five more according to external gluons permutations),  here the Q-representation for the scalar loop has the following bulk terms:

\begin{figure}[h]
%\begin{picture}(0,0)(6000,10000)
\centering
\includegraphics[width=2.5in]{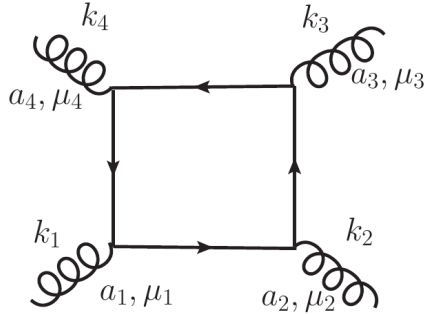}
\caption{Four-gluon vertex.}
\label{fig2}
\end{figure}

\bear
\Gamma^{a_1a_2a_3a_4} &=&  
g^4\mbox{tr}(T^{a_{1}}\dots T^{a_{4}})
 %(2\pi)^{D}i\delta(\sum p_{i})
 \int_{0}^{\infty} dT(4\pi T)^{-D/2}e^{-m^2 T}\nonumber\\
 && \times\int_{0}^{T}d\tau_{1}\int_0^{\tau_{1}}d\tau_2\int_0^{\tau_2}d\tau_3
 Q_4
\exp\bigg\{\sum_{i,j=1}^4\frac{1}{2} G_{Bij}k_{i}\cdot k_{j}
\bigg\}\,,
\nonumber\\
\ear
where
\bear
Q_4&=&Q^4_4+Q^3_4+Q^2_4-Q^{22}_{4}\,,\non\\
Q^4_4&=&\Gd(1234)+\Gd(1243)+\Gd(1324)\,,\non\\
Q^3_4&=&\Gd(123)T(4)+\Gd(234)T(1)+\Gd(341)T(2)+\Gd(412)T(3)\,,\non\\
Q^2_4&=&\Gd(12)T(34)+\Gd(13)T(24)+\Gd(14)T(23)+\Gd(23)T(14)\non\\
&&+\Gd(24)T(13)+\Gd(34)T(12)\,,\non\\
Q^{22}_4&=&\Gd(12)\Gd(34)+\Gd(13)\Gd(24)+\Gd(14)\Gd(23)\,,\nonumber\\
\ear
and we have now employed a more condensed notation:

\bear
\Gd(i_1i_2\cdots i_n)&:=&\Gd_{Bi_1i_2}\Gd_{Bi_2i_3}\cdots\Gd_{Bi_ni_1}\Bigl(\half\Bigr)^{\delta_{n,2}} {\rm tr}(f_{i_1}f_{i_2}\cdots f_{i_n})\,,\nonumber\\
T(i)&:=&\sum_r\Gd_{Bir}\epsk ir\,,\nonumber\\
T(ij)&:=&\sum_{r,s}\Bigg\{\Gd_{Bir}\epsk ir\Gd_{js}\epsk js+\half\Gd_{Bij}\epseps ij\Big\lbrack\Gd_{Bir}\kk ir-\Gd_{Bjr}\kk jr\Big\rbrack\Bigg\}\,.\nonumber\\
\ear
The IBP procedure now leads to both single boundary terms (three-point integrals) and double boundary terms (two-point integrals).
The following rules emerge:

\begin{itemize}

\item
Each single boundary term, say for the limit $ 3\to 4$, matches  some bulk term in the Q-representation of the three-gluon vertex,
with momenta $(k_1,k_2,k_3+k_4)$, and $f_3=k_3 \otimes \varepsilon_3- \varepsilon_3\otimes k_3$ 
replaced by $\varepsilon_3\otimes \varepsilon_4 -  \varepsilon_4\otimes \varepsilon_3 $.

\item
Each double boundary term, say for the limit $1\to2 , 3\to 4$, matches  the bulk term in the Q-representation of the 
two-point function,
with momenta $(k_1+k_2,k_3+k_4)$, and the double replacement
\bear
f_1& =&k_1 \otimes \varepsilon_1- \varepsilon_1\otimes k_1
 \to 
 \varepsilon_1\otimes \varepsilon_2 -  \varepsilon_2\otimes \varepsilon_1 \,,
\nonumber\\
 f_2 &= &k_2 \otimes \varepsilon_2- \varepsilon_2\otimes k_2
 \to 
 \varepsilon_3\otimes \varepsilon_4 -  \varepsilon_4\otimes \varepsilon_3 \,.
\nonumber\\
\ear
\end{itemize}
Moreover, this recursive structure is compatible with the replacement rules.

The S-representation looks similar, but has the bulk terms written completely in terms of the $f_i$, so that all non-transversality has now been absorbed into the boundary terms. 

\section{Tensor decomposition of the four-gluon vertex}
\label{sec:Tlist}

Up to permutations compatible with the fixed color-ordering (that is, cyclic permutations and inversion) 
at the one-loop level 19 different structures appear in our representation. This is independent of the
spin in the loop. However, as explained above of these 19 tensors five are just the two- and three-point form factors reappearing at the
four-point level as boundary terms. Let us list here the remaining 14 ``true'' four-point tensors:
\bear
T^4_P & = & \tr (f_1f_2f_3f_4)\,, \nonumber\\
T^4_{NP} & = & \tr (f_1f_3f_2f_4)\,, \nonumber\\
T^{22}_P &=& \fourth \tr (f_1f_2)\tr (f_3f_4)\,, \nonumber\\
T^{22}_{NP} &=& \fourth \tr (f_1f_3)\tr (f_2f_4)\,, \nonumber\\
T^{3}_P &=& \tr (f_1f_2f_3)\veps_4\cdot k_1\,, \nonumber\\ 
T^{3}_{NP} &=& \tr (f_1f_2f_3)\veps_4\cdot k_2\,, \nonumber\\ 
T^{2adj}_{\rm quart} &=& \half \tr (f_1f_2) \veps_3\cdot k_1 \veps_4\cdot k_1\,, \nonumber\\
T^{2adj}_P &=& \half \tr (f_1f_2) \veps_3\cdot k_2 \veps_4\cdot k_1\,, \nonumber\\
T^{2adj}_{NP} &=& \half \tr (f_1f_2) \veps_3\cdot k_1 \veps_4\cdot k_2\,, \nonumber\\
T^{2adj}_C &=& \half \tr (f_1f_2) \Bigl(\veps_3\cdot k_4 \veps_4\cdot k_1 -\half \veps_3\cdot \veps_4 k_4\cdot k_1\Bigr)\,, \nonumber\\
T^{2adj}_Z &=& \half \tr (f_1f_2) \Bigl(\veps_3\cdot k_4 \veps_4\cdot k_2 -\half \veps_3\cdot \veps_4 k_4\cdot k_2\Bigr)\,,  \nonumber\\
T^{2opp}_{\rm quart} &=& \half \tr (f_1f_3) \veps_2\cdot k_1 \veps_4\cdot k_1\,, \nonumber\\
T^{2opp}_P &=& \half \tr (f_1f_3) \veps_2\cdot k_3 \veps_4\cdot k_1\,, \nonumber\\
T^{2opp}_{NP} &=& \half \tr (f_1f_3) \Bigl(\veps_2\cdot k_4 \veps_4\cdot k_1- \half \veps_2\cdot \veps_4 k_4\cdot k_1\Bigr)\,. \nonumber\\
\label{Tlist}
\ear
Here the upper indices $4,3,2,22$ as before refer to the cycle structure. $`adj'$ refers to $f_1$ and $f_2$ being adjacent on the loop, 
$`opp'$ to $f_1$ and $f_3$ being opposite on the loop in the standard ordering. The lower indices refer to the shape of the
corresponding `worldline Feynman diagrams', which to explain here would lead us too far.

\section{List of one-loop Schwinger parameter integrals}
\label{sec:Plist}

At the one-loop level and for a scalar loop, each of the 14 structures (\ref{Tlist}) contributes to the (color-ordered,
dimensionally continued) amplitude as follows:

\bear
\Gamma_{0,l}^{a_1a_2a_3a_4,u} =
 \frac{g^4}{(4\pi)^{\frac{D}{2}}}(T^{a_1}T^{a_2}T^{a_3}T^{a_4})\Gamma\Big(4-\frac{D}{2}\Big)T_{l}^{u}
\int_0^1\prod_{i=1}^4d\alpha_i\delta\bigl(1-\sum_{i=1}^4\alpha_i\bigr)\frac{P_{0,l}^u}
{{\rm Den}^{4-\frac{D}{2}}}\, .\nonumber\\
\ear
Here 
\bear
{\rm Den}  \equiv m^2+\alpha_1\alpha_2k_1^2+\alpha_2\alpha_3k_2^2+\alpha_3\alpha_4k_3^2+\alpha_1\alpha_4k_4^2+\alpha_1\alpha_3(k_1+k_2)^2+\alpha_2\alpha_4 (k_1+k_3)^2\,,
\nonumber\\
\label{DefDen}
\ear
is the standard four-point off-shell denominator polynomial, and below we list the 14 numerator polynomials:

\bear
P^4_{0,P}&=&(1-2\alpha_1)(1-2\alpha_2)(1-2\alpha_3)(1-2\alpha_4)\,,\nonumber\\
P^4_{0,NP}&=&-(1-2\alpha_1)(1-2\alpha_3)(1-2\alpha_2-2\alpha_3)(1-2\alpha_3-2\alpha_4)\,,\nonumber\\
P^{22}_{0,P}&=&(1-2\alpha_2)^2(1-2\alpha_4)^2\,,\nonumber\\
P^{22}_{0,NP}&=&(1-2\alpha_2-2\alpha_3)^2(1-2\alpha_3-2\alpha_4)^2\,,\nonumber\\
P^3_{0,P}&=&-(1-2\alpha_1)(1-2\alpha_2)(1-2\alpha_3)(1-2\alpha_2-2\alpha_3)\,,\nonumber\\
P^3_{0,NP}&=&(1-2\alpha_2)(1-2\alpha_3)(1-2\alpha_2-2\alpha_3)(1-2\alpha_3-2\alpha_4)\,,\nonumber\\
P^{2adj}_{\rm 0, quart}&=&(1-2\alpha_1)(1-2\alpha_2-2\alpha_3)(1-2\alpha_2)^2\,,\nonumber\\
P^{2adj}_{0,P}&=&(1-2\alpha_1)(1-2\alpha_3)(1-2\alpha_2)^2\,,\nonumber\\
P^{2adj}_{0,NP}&=&-(1-2\alpha_2-2\alpha_3)(1-2\alpha_3-2\alpha_4)(1-2\alpha_2)^2\,,\nonumber\\
P^{2adj}_{0,C}&=&-(1-2\alpha_1)(1-2\alpha_4)(1-2\alpha_2)^2\,,\nonumber\\
P^{2adj}_{0,Z}&=&(1-2\alpha_4)(1-2\alpha_3-2\alpha_4)(1-2\alpha_2)^2\,,\nonumber\\
P^{2opp}_{\rm 0,quart}&=&(1-2\alpha_2)(1-2\alpha_1)(1-2\alpha_2-2\alpha_3)^2\,,\nonumber\\
P^{2opp}_{0,P}&=&-(1-2\alpha_1)(1-2\alpha_3)(1-2\alpha_2-2\alpha_3)^2\,,\nonumber\\
P^{2opp}_{0,NP}&=&-(1-2\alpha_1)(1-2\alpha_3-2\alpha_4)(1-2\alpha_2-2\alpha_3)^2\,.\nonumber\\
\ear 
These bulk contributions are UV finite in $D=4$. 

As a check on our four-gluon results we have used them for a complete recalculation of the
low-energy effective action for the scalar loop case, and found perfect agreement with \cite{vandeven,25}.

\section{Conclusions and outlook}
\label{sec:conc}

\no
To summarize, the main points which we wanted to make here are:

\begin{itemize}

\item
In the string-inspired formalism, form factor decompositions of the $N$ - vertex compatible with Bose symmetry
and gauge invariance can be generated simply by an integration-by-parts procedure
starting from the Bern-Kosower master formalism, which originally was derived as a generating functional
for on-shell matrix elements. 

\item
At the one-loop level, the parameter integrals appearing in the form factors for the scalar, spinor and gluon
loop cases are all obtained directly from the Bern-Kosower master formula.

\item
We have carried out this program explicitly for the three- and four-point cases. 

\item
In particular, we have obtained a natural four-point generalization of the Ball-Chiu
decomposition. It is distinguished by the fact that all true four-point terms are manifestly transversal,
so that all longitudinal components are given by lower-point integrals. It contains only 19 
different tensors structures, of which only 14 have full four-point kinematics. 

\end{itemize}

The compactness of our representation should make it very useful, for example, as an input for the
DS equations (in this context, previously only the two and three-point
amplitudes were used with their full loop-corrected structure, but a need for the inclusion of the four-gluon
vertex is already felt \cite{kelfis,madalk}).

Our results also underline the importance of further developing the string-inspired worldline formalism,
e.g. to the case of amplitudes involving open scalar lines \cite{abc,102}, as well as the original string-based 
Bern-Kosower approach (see \cite{cornwall,magnea} for recent progress in this direction). 

\section*{Acknowledgements}
We would like to thank A. Bashir, W. Bietenholz, M. Cornwall, P. Dall'Olio, A. Davydychev, L. Dixon, G. Eichmann, C. Fischer, J. Gracey, P. Hess, L. Magnea, J. Papavassiliou, 
and A. Weber for helpful discussions and correspondence. C. S. thanks CONACYT for financial support through grant CB2014 242461. 
The work of N. A. was supported by IBS (Institute for Basic Science) under grant IBS-R012-D1.

\begin{appendix}

\section{Summary of Conventions}
\label{conv}

\renewcommand{\theequation}{A.\arabic{equation}}
\setcounter{equation}{0}
\vskip10pt

We work with the $(-+++)$ metric. 
The nonabelian covariant derivative is
$D_\mu\equiv \partial_\mu+ig A^a_\mu  T^a$,
with $[T^a,T^b] = i f^{abc}T^c$. The adjoint
representation is given by $(T^a)^{bc}=-i f^{abc}$.
The normalization of the generators is 
${\rm tr}(T^aT^b) = C(r) \delta^{ab}$, where
for $SU(N)$ one has $C(N)=\frac{1}{2}$ for 
the fundamental and $C(G)= N$ for the adjoint representation. 

\end{appendix}

\end{document}